\documentclass[aps,reprint,nofootinbib,showkeys]{revtex4-2}

\usepackage{epsfig}
\usepackage{amsmath}
 \usepackage{hyperref}
\usepackage{microtype}
\usepackage[usenames]{xcolor}
\usepackage[lined,boxed,commentsnumbered]{algorithm2e}

\usepackage{graphicx,natbib}
\usepackage{amssymb}

\def\be{\begin{equation} }
\def\ee{\end{equation} }

\def\bea{\begin{eqnarray} }
\def\eea{\end{eqnarray} }

\def\beas{\begin{eqnarray*} }
\def\eeas{\end{eqnarray*} }

\newtheorem{example}{Example}{}

\newtheorem{definition}{Definition}{}

\newtheorem{theorem}{Theorem}{}

\newtheorem{conjecture}{Conjecture}{}

\RequirePackage{natbib}

\begin{document}
\title[An out-of-equilibrium 1D particle system]{An Out-of-Equilibrium 1D Particle System Undergoing Perfectly Plastic Collisions}

\author{Daniel Fraiman $^{\dag\S}$}

\email{dfraiman@udesa.edu.ar}

  \affiliation{$^\dag$ Departamento de Matem\'atica y Ciencias, Universidad de San Andr\'es, Argentina,}
  \affiliation{$^\S$ Conicet, Argentina.}

\begin{abstract}
At time zero, there are $N$ identical point particles in the line (1D) which are characterized by their positions and velocities. Both values are given randomly and independently from each other, with arbitrary probability densities. Each particle evolves at constant velocity until eventually they meet.  When this happens, a perfectly-plastic collision is produced,  resulting in a new particle composed by the sum of their masses and the weighted average velocity.
 The merged particles evolve indistinguishably from the non-merged ones, i.e. they move at constant velocity until a new plastic collision eventually happens.
 As in any open system, the particles are not confined to any region or reservoir, so as time progresses, they go on to infinity.
 From this non-equilibrium process, the number of (now, non-identical) final particles, $\tilde{X}_N$, the distribution of masses of these final particles and the kinetic energy loss from all plastic collisions, is studied. The principal findings shown in this paper are outlined as follows: (1) A method has been developed to determine the number and mass of the final particles based solely on the initial conditions, eliminating the need to evolve the particle system. (2) A similar model of merging particles, with a universal number of final particles, $\tilde{Z}_N$, is introduced. (3) Strong evidence that $\tilde{X}_N$ is also universal and has the same law of probability as $\tilde{Z}_N$ is presented.   (4) An accurate approximation of the energy loss is presented. (5) Results for $\tilde{X}_N$ for an explosive-like initial condition are analyzed.
\end{abstract}
\keywords{Many-body dynamics, out-of-equilibrium systems, 1D system}

\maketitle

\section{introduction}
The significance of non-equilibrium phenomena in physics is profound, as they capture the intrinsic dynamic nature of complex systems beyond their states of thermodynamic equilibrium. While it can be argued that nearly every observable macroscopic event occurs under non-equilibrium conditions, a comprehensive framework for understanding such systems remains elusive. This challenge arises from the diverse array of non-equilibrium phenomena observed in nature. Examples include biological processes~\cite{fang}, chemical systems~\cite{van}, turbulent flows~\cite{pope}, quantum transport in novel materials~\cite{landi}, vehicular movement on road networks~\cite{khairnar,kozlov}, competitive dynamics among populations for resources~\cite{strogatz}, plasma instabilities~\cite{plasma}, among others. Most notably, these phenomena manifest across scales, ranging from the microscopic~\cite{micro} to the cosmological~\cite{cosmo}.

In the study of non-equilibrium phenomena,  complex cases are typically addressed once simpler or more streamlined versions have been established.  In this paper, a very simple system is introduced: a gas consisting of identical point particles in an open one-dimensional space undergoing perfectly plastic collisions. This particular system does not seem to have been rigorously studied before. The results presented here may provide insights into more complex non-equilibrium processes. In particular, the study of non-equilibrium interacting particle systems, such as the one studied here,  could provide valuable insights into the intricate astrophysical phenomena that govern the behavior of stars and galaxies on cosmic scales.

\section{The 1D particle system}
At time zero, there are $N$ identical point particles of mass $m$ in a one-dimensional space at different arbitrary positions. Let the
farthest left particle be considered particle 1, the second be particle 2, and so on, with particle $N$  being the rightmost one, i.e.  their initial positions verify $Y_1<Y_2<\dots <Y_N$ respectively.
 The reason behind the positions distribution being flexible is that these values will not be relevant on what will be studied in this paper. Instead, velocity will take a protagonist role. The initial velocities of each particle $V_1, V_2, \dots, V_N$ are considered as a sequence of iid random variables with an absolute continuous distribution function $F(x):=\mathbb{P}(V_1\leq x)$. Each particle evolves at constant a velocity, $Y_i(t)=V_i t+Y_i$, until it eventually collides with another particle. At this point, a perfectly plastic collision is generated,  resulting in a single particle with a mass that is equal to the sum of the individual particles' masses, which moves at a velocity determined by the conservation of momentum. The conservation of momentum dictates that the velocity of the particle after collision will be the weighted average of the velocities of the particles prior to collision. The merged particles evolve equally to the non-merged ones, i.e. they move at constant velocity until a new plastic collision eventually happens.

In this work, the asymptotic properties of the stochastic process $X_N(t)=\mbox{\textit{number of particles at time} $t$},$
which starts with $X_N(0)=N$ particles are studied. $X_N(t)$ converges to a random variable $\tilde{X}_N$ that naturally depends on $N$,
$$X_N(t) \underset{t \rightarrow \infty}{\rightarrow} \tilde{X}_N.$$
 On the left panel of Fig. 1, a realization of a system of $N=10$ identical particles undergoing perfectly plastic collisions is shown. The evolution of the number of particles, $X_{10}(t)$, is shown below. For this random realization, after a certain amount of time (the last collision time), henceforth referred as $t^{\star}$, the system remains with 3 particles left.  In other words, for this realization, $\tilde{X}_{10}=3$ (or equivalent $X_{10}(t)=3$ for $t\geq t^{\star}$).
 When looking at the mass of these three resulting particles and comparing them to the mass of the initial particles, one of them will have three times the original value; another, six times the original value; and the last one, the one which has not collided with any particles, will have the same mass of the original particles. In other words, the final masses are $\mathbb{M}=(1,6,3)$, with a representation that considers the first coordinate of this vector to be the particle in the farthest left position, the second the next, and so on.

\begin{figure}
  \centering
  \includegraphics[width=8.5cm]{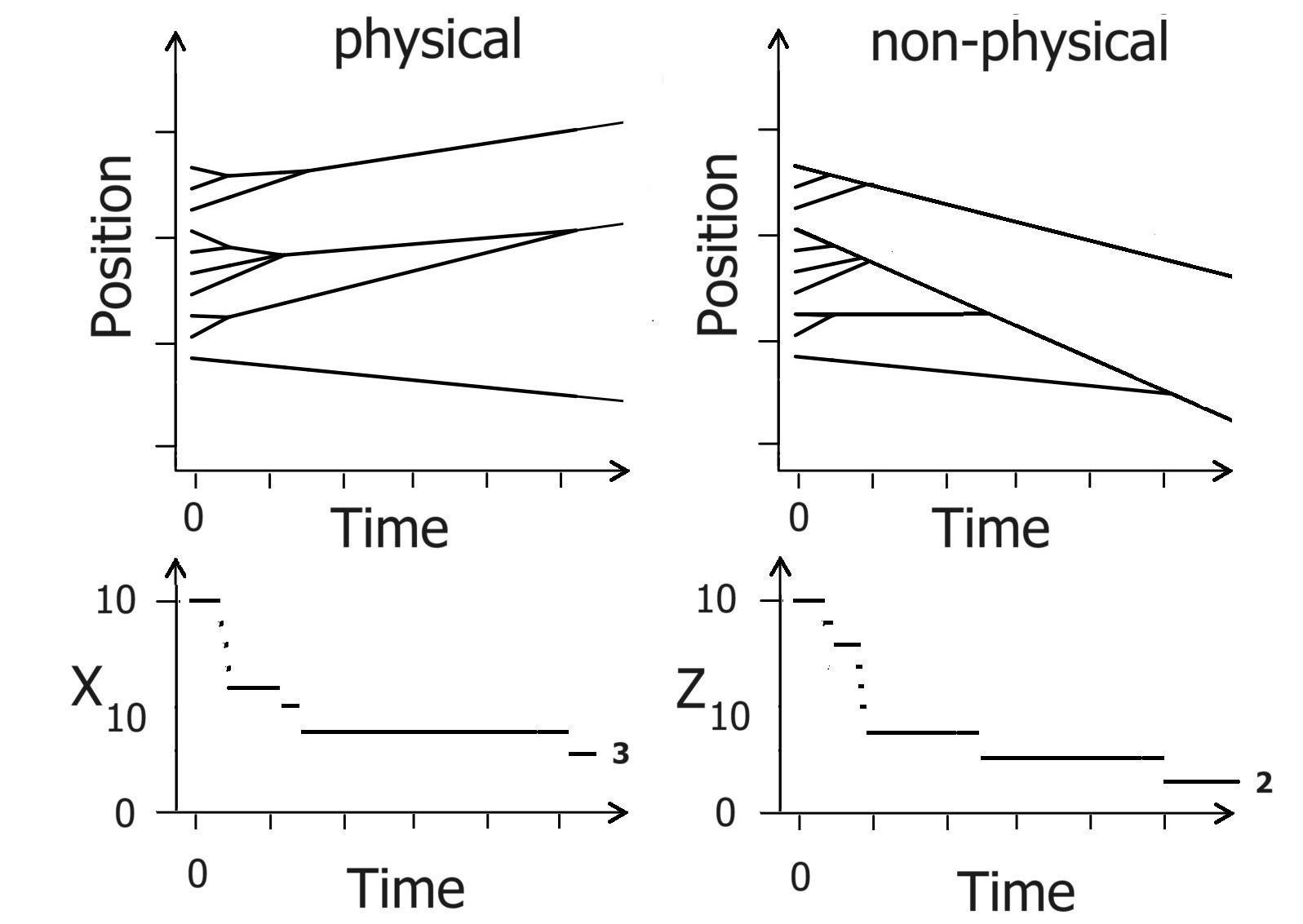}
\caption{Example of the evolution of a system of $N=10$ particles suffering:  perfectly plastic collisions, and non-physical fusion collisions.
 }\label{fig0}
\end{figure}

As one might expect, calculating $\tilde{X}_N$ and $\mathbb{M}$ in this physical model is quite difficult. The strategy for tackling this problem is to introduce a simpler, non-physical model, solve it and apply what has been learned from the simpler model to the physical model. The non-physical model is very similar to the original model, except that when a collision occurs, the velocity of the resulting merged particle is given by the minimum velocity of the two colliding particles before the collision.

With all of this in mind, this paper will be organized in the following manner: the first section will be dedicated to the non-physical model; the second section, to the physical model; and finally,  a conclusion will be introduced, summarizing and discussing the possible changes in the results when a larger dimension is used instead of one dimension (1D), as well as this study's potential for a better understanding of galaxy formation.

\section*{Non-physical model}

A new non-physical collision process for the same initial conditions is developed. In this artificial process, once two particles collide, the resulting merged particle continues at a velocity equal to the minimum velocity of both particles involved in the collision. The only difference between this new process and the original one is that there is no conservation of momentum: the velocity of the merged particle, which is the average velocity of the original particles prior to the merger (see eq.~\ref{velo}), is replaced by the minimum velocity of the colliding particles.  In this new process, considering
$Z_N(t)=\mbox{\textit{number of particles at time} $t$}$,  the number of final particles, now called $\tilde{Z}_N$ is studied again.
 $$Z_N(t) \underset{t \rightarrow \infty}{\rightarrow} \tilde{Z}_N.$$
A realization of this non-physical case is shown in the right panel of Figure 1. The initial conditions are the same as the ones in the physical system shown in the left panel. Note that in this particular realization, $\tilde{Z}_{10}$ is equal to 2 while $\tilde{X}_{10}$ is equal to 3. The evolution of the total number of particles, $Z_{10}(t)$, is shown in the lower right panel of Fig. 1.
 The pattern for calculating the number of final particles can be easily found by looking at this figure. First, the particle with the minimum velocity must be found, let us say it is particle $j$. Then, particles $1,2,\dots,j-1$ merge with particle $j$, producing the leftmost final particle with mass $j$. Then, the minimum velocity particle among the remaining particles $j+1,j+2,\dots,N$ is identified,  let us say it is particle $k$. Then,  all the previous particles in this group ($j+1,j+2,\dots,k$) are merged, producing a second final particle of mass $k-j$. This process is repeated until the minimum velocity particle is particle $N$.  For example, in the right panel of Fig. 1, it can be observed that the minimum velocity corresponds to particle number 7, which merges particles from 1 to 7, and the second minimum velocity of the remaining particles corresponds to particle 10, which merges particles from 8 to 10. So the final two particles ($\tilde{Z}_{10}=2$) have masses of 7 and 3 respectively (defined as $\mathbb{K}=(7,3)$ to differentiate $\mathbb{M}$ from of the physical model). It is important to note that this final configuration remains the same even if the initial positions are modified, however, their order can not be altered. I.e. $\tilde{Z}_{10}=2$ and $\mathbb{K}=(7,3)$ for the given initial velocities and for any initial position satisfying $Y_1< Y_2< \dots <Y_{10}$. This will always be the case; the positions will be irrelevant, and the only thing that matters will be the order of the particle velocities, not the specific values. That is why the behavior of $\tilde{Z}_{10}$ and their masses are described as universal or distribution-free. Below are some definitions related to what has been discussed and the main result.

Let $\Theta_N=\{1,2,\dots,N\}$ and $\Omega_N=\{\omega_1,\omega_2,\dots,\omega_{N!}\}$ be the set containing the $N!$ sequences that can be formed with all $N$ elements of $\Theta_N$ without replacement, and let $\Xi=\{\xi_1,\xi_2,\dots,\xi_{m}\}$ with $m=\sum_{i=1}^{N-1}\binom{N}{N-i}(N-i)!$ represent all the sequences of size smaller than $N$ that can be formed with the elements of $\Theta_N$ without replacement. The space in which all these sequences live will be called $\Sigma_N:=\Omega_N \cup \Xi$. Frow this point forward, all sequences will be considered vectors. Let $s$ be an arbitrary sequence, $s[j]$ correspond to element $j$, and $s[j:k]$ correspond to the subsequence starting at coordinate $j$ and ending at element $k$ of the sequence.
\begin{definition}
Let $s\in \Sigma_N$ be an arbitrary sequence, $L(s)$ be the length of the sequence, and the coordinate or place of the sequence with the minimum value as
\begin{equation*}
\tilde{k}(s)=\{k\in \Theta_{L(s)}: s[k]=min(s)\},
\end{equation*}
and  $\tilde{k}(\emptyset)=L(\emptyset)=0$.
\end{definition}

\begin{definition}
Let $Q$ be a function $Q:\Sigma_N \cup \emptyset  \to \Sigma_N \cup \emptyset$, called here cut function, that verifies
\begin{equation*}
Q(s)=\left\{
\begin{array}{lll}
s[(\tilde{k}(s)+1):L(s)] &  &   \mbox{if}\ \ L(s) > \tilde{k}(s)  \\
\emptyset &  & \mbox{if}\ \     L(s) = \tilde{k}(s).
\end{array}
\right.
\end{equation*}
The last condition includes $Q(\emptyset)=\emptyset$.
\end{definition}

\begin{definition}
Let $Q^k$ be the k-times composition of the function $Q$, with $Q^0$ the identity function.
For example, $Q^3(s)=Q(Q(Q(s)))$. The integer function $\tilde{Z}_N: \Omega_N \to \Theta_N $ is defined as
\begin{equation}
\tilde{Z}_N(s)=\min \{k\in \Theta_{N}: Q^k(s)=\emptyset\}.
\end{equation}
\end{definition}

\begin{definition}
For a given sequence, the cluster array $\mathbb{K}(s)$ is defined as,
\begin{equation*}
\mathbb{K}(s)=(\tilde{k}(Q^0(s)),\tilde{k}(Q^1(s)),\tilde{k}(Q^2(s)),\dots, \tilde{k}(Q^{\tilde{Z}_N(s)-1}(s))).
\end{equation*}
\end{definition}

\begin{example}
Let $s=(5,2,8,1,9,3,10,7,4,6)$, then:
\[ \begin{array}{lcl}
  \tilde{k}(s)=4 & &Q^1(s)=(9,3,10,7,4,6), \\
  \tilde{k}(Q^1(s))=2 & & Q^2(s)=(10,7,4,6), \\
  \tilde{k}(Q^2(s))=3 & & Q^3(s)=(6),  \\
  \tilde{k}(Q^3(s))=1 & & Q^4(s)=\emptyset,  \end{array}\]
 $\tilde{Z}_{10}(s)=4$ and $\mathbb{K}(s)=(4,2,3,1)$.
\end{example}

Most importantly in this section, if $w$ is considered a randomly chosen sequence $\omega \in \Omega_N$, then $$\tilde{Z}_N=\tilde{Z}_N(w).$$
 Furthermore, the probability that $\tilde{Z}_N$ takes the value $k$ with $k\in \Theta_N$ can be calculated as follows
\begin{equation}\label{proba}
 \mathbb{P}(\tilde{Z}_N=k)=\frac{|\{\omega \in \Omega_{N}: \tilde{Z}_N(\omega)=k\}|}{|\Omega_{N}|}.
\end{equation}
As shown in eq.~\ref{proba},  the problem of calculating the punctual probability of $\tilde{Z}_N$ is a combinatorial one, e.g. $\mathbb{P}(\tilde{Z}_N=N)=\frac{1}{N!}$. The following theorem is presented for the general case.
\begin{theorem}
The punctual probability of $\tilde{Z}_N$ verify,
\begin{equation}\label{distri}
\mathbb{P}(\tilde{Z}_N=k)=\frac{|c(N,k)|}{N!}
\end{equation}
where $c(n,k)$ is the Stirling number of the first kind given by the equality
$a(a - 1)\dots(a - N + 1) = \overset{N}{\underset{k=0}{\sum}} c(N, k)a^k$.
\end{theorem}
For the proof, it is sufficient to note that if the initial velocities of particles $V_1, V_2, \dots, V_N$ are considered to be a time series. As usual, the index corresponding to velocities represents discrete time. The problem of finding the number of records of minimum value in this time series has punctual probability given by eq.~\ref{distri}. In other words, the problem of finding how many record values (ocurrence of maximum/minimium value) there are in a time series generated by a sequence of iid continuous random variables, is equivalent to finding the number of final particles in the non-physical model.  Remarkably, the problem of finding the number of record values can be traced back to Renyi~\cite{Renyi,Renyi2}. See~\cite{records1,records2} for additional references on record analysis.

 In addition, the mean and variance of  $\tilde{Z}_N$ verify,
\begin{equation}\label{meanf}
 \langle \tilde{Z}_N \rangle =\overset{N}{\underset{k=1}{\sum}}  \frac{1}{k},
\end{equation}
\begin{equation}\label{varf}
 \langle \tilde{Z}^2_N \rangle-\langle \tilde{Z}_N \rangle^2 =\overset{N}{\underset{k=1}{\sum}} \ \frac{1}{k}-\overset{N}{\underset{k=1}{\sum}} \ \frac{1}{k^2}.
\end{equation}
 Note that for large $N$, the previous equations are represented by
\begin{equation}\label{scaling}
\langle \tilde{Z}_N \rangle  \approx \ln(N)+\gamma,
\end{equation}
\begin{equation}\label{scaling2}
\langle \tilde{Z}^2_N \rangle-\langle \tilde{Z}_N \rangle^2  \approx \ln(N)+\gamma -\frac{\pi^2}{6},
\end{equation}
where $\gamma$ is the Euler-Mascheroni constant.

It is important to stress that the results presented here are universal (or distribution free). I.e. equation 3 (and the ones derived from it, eqs.4-7) is valid for any initial position distribution that preserves the order and for any continuous initial velocity distribution.  In essence, the specific distribution of initial particle positions and velocities does not matter. This behavior is expected in systems approaching a critical phase transition~\cite{critico1,critico2} or in self-organized criticality systems~\cite{soc1,soc2,soc3,soc4,soc5,soc6}.

In summary, for the non-physical system introduced in this paper, the behaviour of the final number of particles has been found to be universal. Punctual probability and the first two moments have been explicitly calculated. The next step in the following section is to study the physical model through the eyes of the non-physical one.

\section*{Physical model}
In this section, the originally proposed model in which collisions preserve momentum is studied. This section contains:  (A) A procedure for calculating $\tilde{X}_N$ and $\mathbb{M}$ without time evolution, (B) A study of the universal behaviour of $\tilde{X}_N$ and its (C) distribution, mean and variance, (D) A study of the distribution of the masses of each of the $\tilde{X}_N$ final particles, (E) Simulations and an accurate theoretical approximation for the mean fraction of energy loss after all plastic collisions, and (F) Results for an explosive initial condition.

\subsection{A calculus for $\tilde{X}_N$ and $\mathbb{M}$ that does not require time evolution}
For any given initial condition, the way to compute the number of final particles ($\tilde{X}_N$) and their masses has been to evolve the particles system numerically and study the results. However, is there a way to calculate these values without evolving the system?
The main purpose of this section is to show that it is possible, and to present a proposal for calculating both $\tilde{X}_N$ and the mass of each final particle, without time evolution.

Before presenting the main result,  two points must be understood. First, note that in a system characterized by perfectly plastic collisions, any fused particle (regardless of the order in which the collisions occurred)  will have a velocity that is the average of the velocity of the particles that formed it.  Specifically, when two particles with masses $m_1$ and $m_2$ and velocities $V_1$ and $V_2$ respectively collide,  the fused particle of mass $m_1+m_2$ has a velocity equal to
$V_1p_1+V_2(1-p_1)$ with $p_1=m_1/(m_1+m_2)$. Calculating the velocity of a system of particles of equal masses, such as the one in this case, is surprising simple: the final velocity of any fused particle, $\tilde{V}_f$, formed by $N_f$ identical particles, such as particles $k,k+1,\dots,k+N_f-1$,  ends up being the average velocity of the fused particles,
\begin{equation}\label{velo}
\tilde{V}_f=\overline{V}_{k,k+N_f-1},
\end{equation}
where $\overline{V}_{i,j}:=\frac{1}{j-i}\underset{k\in \Theta_{i,j}}{\sum}V_k$ with $\Theta_{i,j}:=\{k \in \mathbb{N}:  i \leq k \leq j \}$.

Another important point to highlight is that, similarly to the non-physical model, changing the initial positions of each particle, while maintaining the order of the particles, does not alter the number of final particles or their masses. These position changes can only affect the sequence of collisions and, trivially, the timing of the last collision,  but they do not affect the final configuration. The theorem below expresses this fact.
 \begin{theorem}
 Let $Y_1< Y_2 < \dots < Y_N$ denote arbitrary initial positions of the $N$ particles, respectively, and let $V_1, V_2, \dots, V_N$ represent their initial velocities. Then $\tilde{X}_N$ and $\mathbb{M}$ do not depend on the initial positions.
 \end{theorem}
  See~\hyperref[appA]{Appendix A} for proof.

Before formally presenting the main result of this section (Theorem 3), let's first provide a simple explanation of this result. The method for determining the final number of particles is analogous to that in the non-physical model. Initially,  a particle that meets a specific condition is identified. Subsequently, the identified particle is merged with the particles with indices lower to it. From the remaining particles, another particle that satisfies the given condition must then be found. Again, this particle is merged with the particles with indices lower to it. This process continues iteratively until particle $N$ satisfies the condition. In the non-physical model, the condition is based on the minimum velocity. However, in the physical model, the condition is more intricate. Specifically, one must find a particle, denoted as particle $k$, whose velocity is such that the resulting merged particle has a velocity lower to any merged particle containing particle $k+1$.

Now, some definitions similar to those introduced for the non-physical model and the main result are presented.
 Consider the vector $v_i$, representing the initial velocities of the $N$ particles, expressed as $v_i = (V_1, V_2, \dots, V_N)$. Let $v$ be a general vector obtained by potentially excluding the first coordinates, such as $v$ = ($V_4, V_5, \dots, V_N$). The set of all real vectors $v$ with lengths ranging from 1 to $N$ will be referred to as $\Lambda := \underset{k\in \Theta_N}{\cup} \mathbb{R}^k$.

\begin{definition}
Let  $\tilde{m} : \Lambda \to \Theta_N$
be the function
 \begin{equation*}
\tilde{m}(v)=\min\{j \in \Theta_{1,L(v)}: \overline{v}_{1,j}<  \overline{v}_{j+1,i} \quad \forall i \in \Theta_{j+1,L(v)+1}\}
\end{equation*}
 where  $\overline{v}_{i,j}=\frac{1}{j-i}\overset{j}{\underset{k=i}{\sum}} w[k]$, with
  \begin{equation*}
w[k]=\left\{
\begin{array}{lll}
v[k] &  &   \mbox{if}\ \ k \in \Theta_{L(v)} \\
2\max\{v[1],v[2], \dots, v[L(v)]\} &  & \mbox{if}\ \ k =L(v)+1.
\end{array}
\right.
\end{equation*}
By hypothesis, the velocities are independent continuous random variables, so there will be no repeated values. Therefore,  $\tilde{m}(v)$ adopts a unique value.
\end{definition}

\begin{definition}
Let $G$ be a function 
 that verifies
\begin{equation*}
G(v)=\left\{
\begin{array}{lll}
v[(\tilde{m}(v)+1):L(v)] &  &   \mbox{if}\ \ L(v) > \tilde{m}(v)  \\
\emptyset &  & \mbox{if}\ \     L(v) = \tilde{m}(v).
\end{array}
\right.
\end{equation*}
The last condition includes $G(\emptyset)=\emptyset$
\end{definition}

\begin{definition}
Let $G^k$ be the k-times composition of the function $G$, with $G^0$ the identity function. For example, $G^3(v)=G(G(G(v)))$.
\end{definition}

\begin{theorem} It is possible to calculate the number of final particles ($\tilde{X}_N$) and their individual masses ($\mathbb{M}$) without evolving the particle system (in time). Moreover, for a given initial velocity vector $v_i=(V_1,V_2,\dots, V_N)$,
\begin{equation}
 \tilde{X}_N(v_i)=min\{k\in \Theta_{N}: G^{k}(v_i)=\emptyset\},
\end{equation}
\begin{equation}
 \mathbb{M}(v)=(\tilde{m}(G^0(v_i)),\tilde{m}(G^1(v_i)),\tilde{m}(G^2(v_i)),\dots, \tilde{m}(G^{\tilde{X}_N(v_i)-1}(v_i))).
 \end{equation}
\end{theorem}
See~\hyperref[appB]{Appendix B} for proof. Based on the previous Theorem, a simple algorithm for computing $\tilde{X}_N$ and $\mathbb{M}$ is presented in~\hyperref[appC]{Appendix C}. Finally, an example is presented.
\begin{example}
Let $v_i=(5.4,2.1,8.5,1.3,9.5,3.7,10.1,7.7,4.6,6.5)$, then:
\[ \begin{array}{lcl}
  \tilde{m}(v_i)=2 & &G^1(v_i)=(8.5,1.3,9.5,3.7,10.1,7.7,4.6,6.5), \\
  \tilde{m}(G^1(v_i))=2 & & G^2(v_i)=(9.5,3.7,10.1,7.7,4.6,6.5), \\
  \tilde{m}(G^2(v_i))=2 & & G^3(v_i)=(10.1,7.7,4.6,6.5),  \\
  \tilde{m}(G^3(v_i))=4 & & G^4(v_i)=\emptyset,  \end{array}\]
    $\tilde{X}_{10}(v_i)=4$, and $\mathbb{M}(v_i)=(2,2,2,4)$.
For completeness, the velocities of the final particles are computed. The velocity of the leftmost merged particle is (5.4+2.1)/2=3.75, and the velocities of the remaining final particles (from left to right) are 4.9, 6.6 and 7.225. Note the increasing velocity behavior of the final particles.
\end{example}

\subsection{Universal behavior of $\tilde{X}_N$}
In this section, evidence in favor of the distribution of $\tilde{X}_N$ being universal is presented, meaning, it does not depend on the distribution of the initial conditions.  Previously, it has been shown that, as long as the order of the particles is preserved, the initial positions are irrelevant in determining the final particle configuration (Thm. 2).
Taking into account that the initial velocities of each particle $V_1, V_2, \dots, V_N$ are considered
a sequence of iid random variables with an absolute continuous distribution function $F(x):= P(V_1 \leq x)$, the question now is: does $F(x)$ affect $\tilde{X}_N$ (and $\mathbb{M}$)? Put in a different way, is $\tilde{X}_N$ (and $\mathbb{M}$) universal (or distribution-free)?

For the non-physical model, the universal behaviour of the number of final particles ($\tilde{Z}_N$) depends only on the order of the velocities and not on the specific values, as explained before.
Here in the physical model, it is natural to assume that the specific values of the velocities are relevant given the condition outlined in Theorem 2 concerning the average velocities of the fusing particles.  It is therefore not evident whether $\tilde{X}_N$ is universal, and it would be natural to assume that it is not. However, this paper argues that it is. In this section, evidence supporting this statement is presented.

The probability law of $\tilde{X}_N$ and its dependence on $F$ is studied here.  The cases of $N=2$ and $N=3$ will be analyzed first.  For $N=2$, the final number of particles ($\tilde{X}_2$) is equal to 2 if and only if $V_1<V_2$, and this occurs with probability 1/2 for any continuous velocity distribution, $\mathbb{P}(\tilde{X}_2=1)=\mathbb{P}(\tilde{X}_2=2)=1/2$.
For $N=3$, $\tilde{X}_3=3$ if $V_1<V_2<V_3$, and this occurs with probability 1/6 for, once more, any continuous velocity distribution. In order to obtain $\tilde{X}_3=2$ the following condition must occur
\begin{equation*}
\begin{split}
& ((V_1>V_2) \cap V_3 > \overline{V}_{1,2}) \cup  ((V_3<V_2) \cap (V_1 < \overline{V}_{2,3}))
\end{split}
\end{equation*}
where
$\overline{V}_{i,j}= (V_i+V_j)/2$. The probability of this event is
\begin{equation*}
\begin{split}
\mathbb{P}&(X_3=2) =\int_0^{\infty}\mathbb{P}(V_1>V_2)\mathbb{P}(V_3>x|\overline{V}_{1,2}=x)g(x)dx\\
& + \int_0^{\infty}\mathbb{P}(V_3<V_2)\mathbb{P}(V_1<x|\overline{V}_{2,3}=x)g(x)dx\\
&=\int_0^{\infty}\frac{1}{2}\mathbb{P}(V_3>x)g(x)dx+\int_0^{\infty}\frac{1}{2}\mathbb{P}(V_1<x)g(x)dx \\
&=\int_0^{\infty}\frac{1}{2}(1-F(x))g(x)dx+\int_0^{\infty}\frac{1}{2}F(x)g(x)dx \\
&=\int_0^{\infty}\frac{1}{2}g(x)dx=\frac{1}{2}.
\end{split}
\end{equation*}
 where $g(x)$ is the probability density of an average of two independent random variables with distribution $F$. Note that once again, $\mathbb{P}(\tilde{X}_3=2)$ (and $\mathbb{P}(\tilde{X}_3=3)$, $\mathbb{P}(\tilde{X}_3=1)$ ) does not depend on $F(x)$. So far, it has been shown that $\tilde{X}_2$ and $\tilde{X}_3$ are universal (distribution-free).

\begin{figure}
  \centering
  \includegraphics[width=6.5cm]{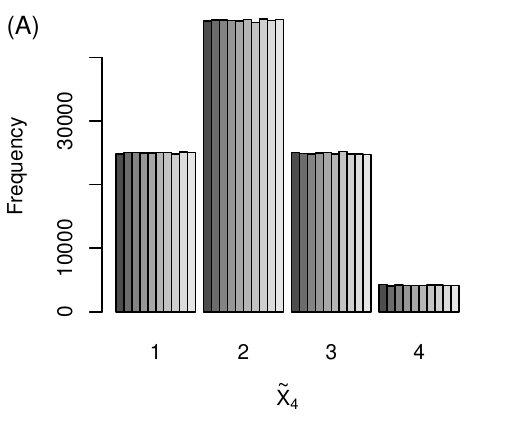}
 \includegraphics[width=6.5cm]{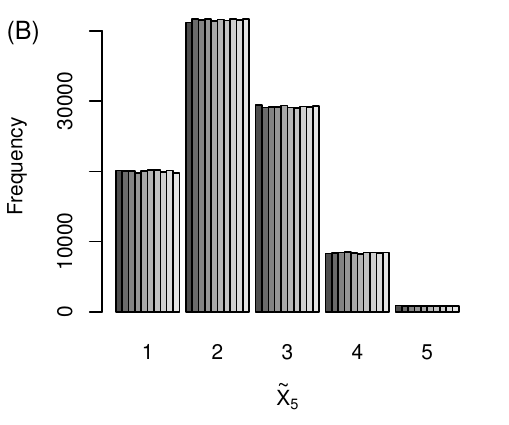}
\caption{Histogram of $\tilde{X}_N$  for ten different velocity distributions, as well as for (A) $N=4$, and (B) $N=5$. Results based on 100000 simulations. The bars from left to right correspond to each of the ten distributions described in the text in the same order as they appear in the text.}
\end{figure}\label{fig_new}

The above calculations become complex even for $N=4$. Therefore, for $N>3$ the results are based on simulations considering different initial velocity distributions:  Uniform(-1,1), Normal(0,1), Normal(10,1), Exponential(1), Exponential(10), Gamma(1,10), Gamma(1,0.1), Beta(1,1), Beta(2,2), Beta($\frac{1}{2}$,$\frac{1}{2}$). The behaviour of $\tilde{X}_N$ is studied by simulations for $N=4,5,6,\dots, 10,100,1000$  and the ten different distributions described above. A comparison of $\tilde{X}_4$ under the different velocity distributions can be found in Fig.~\ref{fig_new}A, and in panel B, the results correspond to $\tilde{X}_5$. Each color bar corresponds to one of the distributions mentioned above. In both cases,  the distribution of the final number of particles is not affected by the initial velocity distribution, as seen in the figure. In fact, when the empirical distributions are statistically compared by the Kolmogorov-Smirnov test adapted to $K$ (greater or equal 2) populations, the equal distribution hypothesis is not rejected at the 0.05 significance level in any of the studied cases (see~\hyperref[apD]{Appendix D} for details of the comparison). Based on the results presented in this section, it is safe to assume that the following conjecture is true.
\begin{conjecture}
For $N\geq 1$, $\tilde{X}_N$ is universal.
\end{conjecture}\label{conjetura1}
Other universal laws for random variables have been discovered over the years. Perhaps one of the most important examples is the first-passage time to the origin for a random walk in discrete time and continuous space. A random walker starts at position zero at time zero, and at time $n$ its position is given by $S_{n}=\overset{n}{\underset{k=1}{\sum}}\psi_k$, where $\psi_1, \psi_2, \dots, \psi_n$ are iid continuous random variables with absolute continuous symmetric distribution $G$. The probability law of the first passage time, $\tau=\min\{n: S_n<0\}$, does not depend on $G$~\cite{andersen,feller,spitzer}. This result is known as the Sparre-Andersen theorem.

\subsection{Distribution, mean and variance of $\tilde{X}_N$}
 The non-physical model, posses an explicit expression for the probability law of $\tilde{Z}_N$ (eq.~\ref{distri}). At this point, it has been conjectured that the random variable number of final particles ($\tilde{X}_N$) is universal (distribution-free). In order to fully understand $\tilde{X}_N$, the best course of action is to obtain an explicit expression for its distribution, just like in the non-physical model. To approach this objective, firstly, the mean and variance of $\tilde{X}_N$ will be studied as a function of $N$, and finally the distribution will be analyzed.

In Section B, explicit expressions for the point probabilities have been obtained for the cases
$N=2$ and $N=3$. Consequently, all moments can be computed for these cases. For $N=2$, the expected value is $\langle \tilde{X}_2 \rangle=1.5$, and the variance is 1/4; and for $N=3$,  $\langle \tilde{X}_3 \rangle=\frac{11}{6}$ and the variance $\frac{17}{36}$. As previously mentioned, the calculations become complex for $N>3$. Therefore, for larger $N$, the results for the mean and variance will be based on simulations using a Uniform(-1,1) distribution for both positions and velocities. According to Conjecture 1, the choice of using a Uniform distribution is irrelevant.

\begin{figure}
  \centering
  \includegraphics[width=8.5cm]{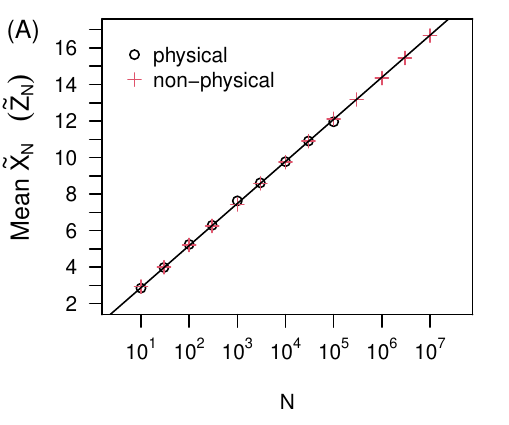}
 \includegraphics[width=8.5cm]{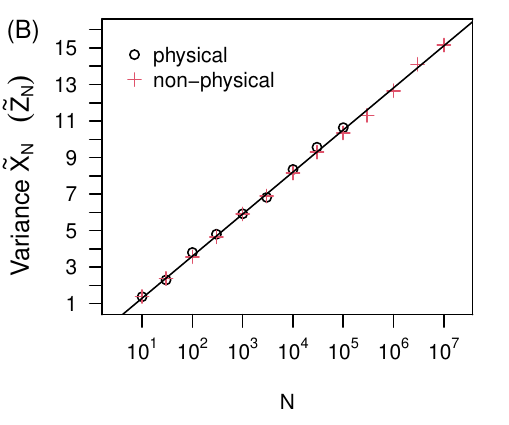}
 \includegraphics[width=8.5cm]{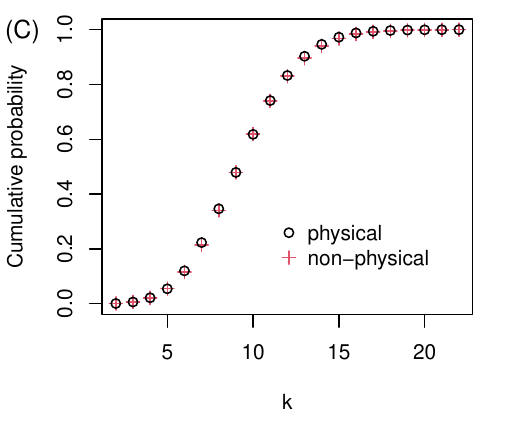}
\caption{(A) Mean and (B) variance of the number of final particles as a function of the number of initial identical particles for both processes. Lines corresponds to (A) $\ln(N)+\gamma$ and (B) $\ln(N)+\gamma-\pi^2/6$.  For each value of $N$ one (ten) thousand realizations were done for the physical (non-physical) system. (C) The empirical probability distribution of the number of final particles for $N=10000$, $\mathbb{P}(\tilde[X]_N<k)$.  The estimation is based on ten thousand realizations for both systems.
 }\label{fig1}
\end{figure}

     The following table shows an estimate of $\langle \tilde{X}_N \rangle$ from numerical simulations, together with the exact values of $\langle \tilde{Z}_N \rangle$ for comparison, for small values of $N=\{2,3,\dots,10\}$.
\begin{table}[h!]
	\begin{center}
		\begin{tabular}{|c|r|l|} \hline
   N & $\langle \tilde{Z}_N \rangle$ \quad  \quad \quad  \quad &\quad  \quad  $\langle \tilde{X}_N \rangle$  \\ \hline
    2  &     $ \frac{3}{2}$\quad \ \ \quad \quad \quad  & \quad
    \quad \quad $\frac{3}{2}$  \\
    3  &    \quad  $ \frac{11}{6}$ \quad \quad \quad \quad  & \quad
    \quad \quad $\frac{11}{6}$  \\
    4  &   $ \frac{50}{24}\approx  2.0833$ &  2.0868 $\pm$  0.0051\\
    5  & $   \frac{274}{120}\approx  2.2833$ &  2.2869 $\pm$ 0.0057 \\
    6  & $  \frac{1764}{720}\approx  2.4500$ &  2.4507 $\pm$ 0.0062 \\
    7  & $ \frac{13068}{5040}\approx  2.5929$ &  2.5911 $\pm$ 0.0066 \\
    8  &$ \frac{109584}{40320}\approx  2.7179$ &  2.7158 $\pm$ 0.0069\\
    9  &$ \frac{1026576}{362880}\approx  2.8290$ &  2.8300  $\pm$ 0.0072 \\
   10 &$ \frac{10628640}{3628800}\approx  2.9290$ &  2.9230 $\pm$ 0.0074 \\ \hline
		\end{tabular}
		\caption{Exact $\langle \tilde{Z}_N \rangle$ and approximate $\langle \tilde{X}_N \rangle$ for different values of $N$. The approximate data corresponds to the 95\% confidence interval obtained from 100000 simulations. For $N=2$ and 3, only exact results are presented.}
	\end{center}
\end{table}
Note that for all values of $N$ shown in Table 1, the confidence interval of $\langle \tilde{X}_N \rangle$ includes the value $\langle \tilde{Z}_N \rangle$, i.e., no evidence of the two values being different was found. Panel A in Fig.~\ref{fig1} provides an estimate of $\langle \tilde{X}_N \rangle$ for larger values of $N$. The results are plotted on a logarithmic scale on the x-axis, and are based on 1000 random realizations for each value of $N$. The expected value shows a linear behavior on this scale, which means that there is a logarithmic behavior. And more importantly, the mean number of final particles in both physical and non-physical models seems to take the same value.

A similar behavior between the physical and the non-physical model is observed for the variance. A table showing that the variance of both $\tilde{X}_N$ and $\tilde{Z}_N$ is indistinguishable is given in~\hyperref[appD]{Appendix D},  and simulations for larger values of $N$ are shown in Fig.~\ref{fig1} B. The variance of $\tilde{X}_N$ has also a logarithm scaling.

Up until now, it has been shown that the first two moments of $\tilde{X}_N$ and $\tilde{Z}_N$ appear to be the same. But what about the remaining moments? If all the moments are equal, this implies that the distributions of $\tilde{X}_N$ and $\tilde{Z}_N$ are identical.

For $N=2$ and 3, both $\tilde{X}_N$ and $\tilde{Z}_N$ distributions are exactly the same. For $N=10000$, Fig.~\ref{fig1}C shows an estimation of the cumulative distribution function of $\tilde{X}_{10000}$, along with the empirical cumulative distribution function for $\tilde{Z}_{10000}$ of the non-physical model. When comparing these two distributions by the Kolmogorov-Smirnov (KS) test, the null hypothesis of equal distribution is not rejected at $5\%$. To further this study,
the distribution of $\tilde{X}_{N}$ for different values of $N$ is studied by running simulations, and compared with the exact distribution of $\tilde{Z}_{N}$ (known by Theorem 1) by the KS test. The results, as shown in~\hyperref[appD]{Appendix D}, indicate that there is no evidence to conclude that the distribution of $\tilde{X}_{N}$ is not the one given by eq.~\ref{distri}. Based on the results presented in this
section, it is safe to assume that the following conjecture is true.
\begin{conjecture}
For $N\geq 1$, $\tilde{X}_N$ and $\tilde{Z}_N$ have the same distribution.
\end{conjecture}\label{conjetura2}
In other words, the conjecture says that the equations \ref{distri}, \ref{meanf}, and \ref{varf} are valid when $\tilde{Z}_N$ is replaced by $\tilde{X}_N$.

\subsection{The mass distribution}
In this section, the mass distribution will be studied numerically. Naturally, one may be tempted to simulate the process, obtain an $\mathbb{M}$ vector, place all the values that compose this vector in a file, create different replicas, read the file, and make a histogram. The problem with doing this is the interpretation: to which random variable does that potential histogram correspond?
To answer this question, let's first note that
$\mathbb{M}$ is a random vector of random length. The fact that the length is not fixed makes the interpretation more difficult and suggests that the strategy outlined above may not be the best one. Therefore, vectors of fixed length will be used and then the combination of them will be performed.

First, the vector $S_N:=\frac{\mathbb{M}}{N}$ is defined: it considers the sizes of the final particles as fractions of the original number of particles. The mass fraction of a randomly chosen final particle, conditioned on the final number of particles being $n$ (and assuming $N$ initial particles) will be denoted as $\tilde{S}_{N,n}$.
 Figure 2 shows the empirical results of $\mathbb{P}(\tilde{S}_{N,n}>s)$ as a function of $s$ for $N=10000$ and $n=[\langle \tilde{X}_{10000} \rangle]=10$ as the rounded value of the expected value of the number of final particles considering eq.~\ref{meanf} and Conjecture 2.  Data of both physical and non-physical processes is presented in a log-$x$ scale. Note that the physical and non-physical processes have the same behavior once more. In the case $n=[\langle \tilde{X}_{N}\rangle]$, it mostly behaves as
 \begin{equation}\label{powerlaw}
 \mathbb{P}(\tilde{S}_{N,n}>s)\approx a_{N}\ln(s),
 \end{equation}
 where $a_{N}$ is a constant that depends on $N$.  For the case under studied in Fig. 2,  $a_N\approx -0.1$. 
 Note that eq.~\ref{powerlaw} is equivalent to say that the probability density function (pdf) of $\tilde{S}_{N,n}$ is a power law function with power exponent equal $1$, i.e.
 $$
  f_{\tilde{S}_{N,n}}(s)\propto \frac{1}{s}  \quad \mbox{for}  \quad \frac{1}{N}\leq s \leq 1.
 $$
 For values of $n$ that differ from $[\langle \tilde{X}_{N}\rangle]$, this logarithmic behavior changes principally at the highest ($1$) and lowest values ($1/N$) of $s$, see~\hyperref[appF]{Appendix F} for details.
\begin{figure}
  \centering
  \includegraphics[width=8.6cm]{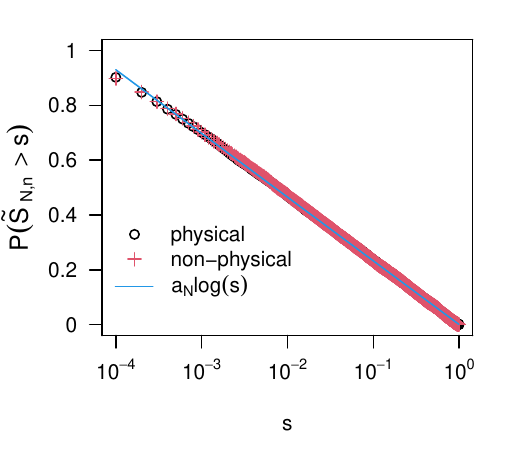}
 \caption{ $\mathbb{P}(\tilde{S}_{N,n}>s)$ as a function of $s$ for $N=10000$ and $n=10$. }\label{fig2}
\end{figure}
Understanding the behavior for different values of $n$ can be difficult; out of the most common distributions, the most consistent with the data shown in Fig.~\ref{fig2} is a Beta, however, confirmation that it is a discrete version of a Beta distribution has not yet been achieved. The behavior of $\mathbb{P}(\tilde{S}_{N}>s)$ is similar to the conditional probability for $n=[\langle \tilde{Z}_{N}\rangle]$, i.e.  $\mathbb{P}(\tilde{S}_{N}>s)\approx  \mathbb{P}(\tilde{S}_{N,[\langle \tilde{Z}_{N}\rangle]}>s)$.

\subsection{An accurate approximation of the kinetic energy loss}

\begin{figure}
  \centering
  \includegraphics[width=8cm]{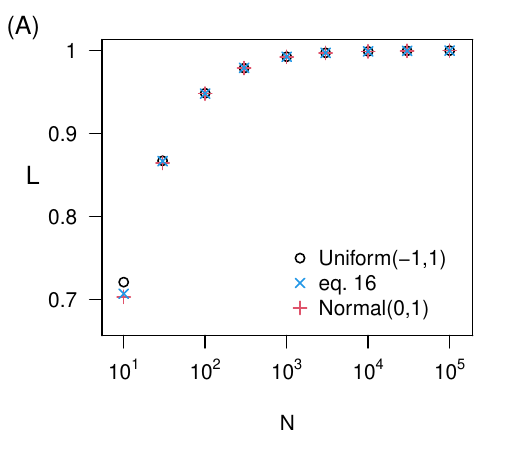}
  \includegraphics[width=8cm]{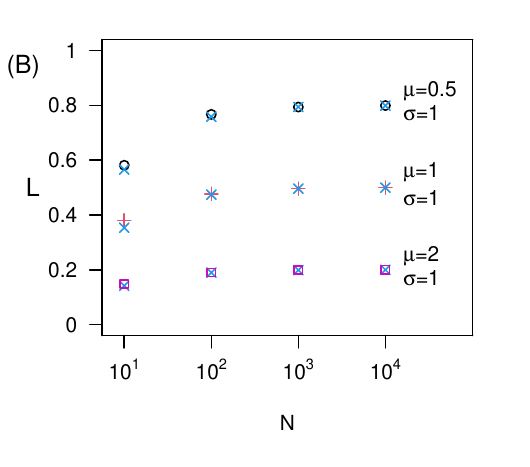}
\caption{Proportion of the initial energy lost by total collisions as a function of the initial number of identical particles,  considering random initial velocities with (A) symmetric distributions: Normal($\mu=0,\sigma=1$) and Uniform(-1,1); and (B) asymmetric  distributions: Normal($\mu=\{0.5,1,2\},\sigma=1$).  The theoretical values of $L$ presented in eq.~\ref{lost} are represented by crosses in both graphs.}\label{fig3}
\end{figure}

In every plastic collision, a fraction of the total energy of the colliding particles is lost. This fraction of energy lost, $L:=(E_{initial}-E_{final})/E_{initial}$, takes values in the interval (0,1). It is equal to 1 in a frontal collision with both having equal absolute momentum values, and is close to 0 in a collision of particles with almost equal velocity. Starting from $N$ particles, the total number of collisions is $N-\tilde{Z}_N$, i.e. it is of the order of the number of particles. Therefore, the amount of energy lost is expected to be large, or, alternatively, a fraction of it is expected to be large. Exactly how large, be it close to 1 or 0.1, is not evident.

Figure 3A shows the average fraction of energy lost,
$\langle L \rangle$,
as a function of system size ($N$).
Given that the initial energy is $E_{initial}=1/2\overset{N}{\underset{i=1}{\sum}}
m V_i^2$, then the final energy will be $E_{final}=\overset{\tilde{X}_N}{\underset{i=1}{\sum}} 1/2 m_i \tilde{V}_i^2$, with $\overset{\tilde{X}_N}{\underset{i=1}{\sum}} m_i=Nm$, where $V_i$ is the initial velocity of the particle $i$, and $\tilde{V}_i$ is the final velocity of the (fused) particle $i$. This figure shows that very similar results are obtained for both Uniform(-1,1) and Normal(0,1) initial velocity distributions. Furthermore, similar results are obtained for other symmetric distributions around zero, such as a double exponential (data not shown).  As seen in the figure, as $N$ increases, the fraction of energy loss increases as well, reaching high levels; over 99\% of energy loss for relatively small systems of 1000 particles. Moreover, $L\to 1$ in the thermodynamic limit.
It is worth emphasizing that, as seen in the observations, the results for small values of N (see N=10 in the graph) are dependent on the velocity distribution. While the velocity distribution had no impact on the previous sections' results, the energy loss, conversely, is influenced by this distribution. Understanding the energy loss calculation requires knowledge of the velocity difference between colliding particles, denoted as $\Delta V:=V_1-V_2$. It's crucial to acknowledge that the distribution of the subtraction of two independent continuous random variables is not universal; each velocity distribution $F$ yields a distinct distribution for $\Delta V$. Therefore, the significance of the velocity distribution is highlighted in this section.

In order to get a theoretical approximation regarding the energy loss, an informal argument about the approximate behaviour of the particle system in a one dimensional space will be explored. Initially,  two ``types of particles'' will be considered: those composed of a large number of merged particles, and those composed of a small number of merged particles, henceforth called $C$ (central) particles and $B$ (border) particles respectively.   The large particles are those that grow linearly with $N$, and the small particles are those that grow sublinearly. The particles that don't start at the ends (i.e. not in the first nor the last positions) will eventually collide and form these large particles ($C$ particles) at the end of the process. On the other hand, particles that start at the ends (or borders) can ``get away'' and suffer very few collisions in this ``escape'', becoming $B$ particles.   The expected velocity of $B$ particles will be similar to the velocity of the original particles since they will suffer few to no collisions. In contrast, the velocity of $C$ particles will be very different from that of $B$ particles. As explained in the previous section, the velocity of fused particles (eq.~\ref{velo}) is an average of independent velocities, meaning that for large fused particles, the velocity will be very close to the initial expected velocity, $\langle V \rangle$. In fact, the variance of the velocity of a fused particle is equal to the variance of the initial velocity divided by the number of particles that form the fused particle ($N_f$), i.e.
$$\langle \tilde{V}_f^2 \rangle=\langle V \rangle^2+\frac{\langle V^2 \rangle-\langle V \rangle^2}{N_f}.$$
The above expression will be used to calculate the velocity of merged $C$ particles. In the case of $B$ particles, the velocity will be approximated by using the initial velocity of the original particle, as if no collision had occurred. Furthermore, a fraction $b$ of the final $\tilde{X}_N$ particles will be considered as particles of type $B$, while the remainder will be treated as merged $C$ particles, with each of them formed by $(N-b\tilde{X}_N)/((1-b)\tilde{X}_N)$ particles.
With all the above considerations, the mean final energy can be written as: %
\begin{equation*}
\begin{split}
& \langle E_{final} \rangle \approx 1/2b \langle \tilde{X}_N\rangle m \langle V^2 \rangle \\
& +1/2(N-b\langle \tilde{X}_N\rangle)m\left(\langle V \rangle^2+\frac{(\langle V^2 \rangle-\langle V \rangle^2)(1-b)\langle \tilde{X}_N\rangle }{N-b\langle \tilde{X}_N\rangle}\right),
\end{split}
\end{equation*}
that does not depend on $b$.
 Therefore, the expected fraction of kinetic energy lost,
\begin{equation}\label{lost}
\langle L \rangle\approx 1- \frac{\langle \tilde{X}_N\rangle}{N} -
(1-\frac{\langle\tilde{X}_N\rangle}{N} ) \frac{\langle V \rangle^2}{\langle V^2\rangle}.
\end{equation}
Note that, for large $N$, this loss of energy $\langle L \rangle $ goes to a value that depends only on the quotient of the first two moments of the velocity distribution. In particular, for velocity distributions with $\langle V \rangle=0$ or $\langle V^2 \rangle=\infty$, all kinetic energy is lost in the thermodynamic limit.

Panel B in Figure 3 is similar to panel A, but in this case, results show for a Normal initial velocity distribution, which is centered at various non-zero values (Normal($\mu=\{0.5,1,2\},\sigma=1$)).   Note that the asymptotic value of $\langle L \rangle$ is now different from 1. For example, the Normal distribution centered at 1 with variance equal to 1 (and $\langle V^2 \rangle=2$) goes to approximately 1/2.

In panels A and B of Fig. 3, the theoretical approximation given by eq.~\ref{lost} is represented by crosses.
 Note that the theoretical approximation for $\langle L \rangle$ is highly accurate in the asymptotic case, and it is also accurate for small values of $N$. Finally, to test the validity of the theoretical approximation given by eq.~\ref{lost}, further velocity distribution were studied by simulating the particle systems for a size of $N=10000$. Table 2 shows the results obtained. Note that the empirical results seem to coincide with the theoretical values in each type of distributions studied, validating our approximation.
\begin{table}[h!]
	\begin{center}
		\begin{tabular}{|c|c|c|} \hline
    Distribution & Empirical  & Theoretical  \\ \hline
Normal(0,1) & 0.99901 $\pm$ 0.00003 & 0.99902 \\
Beta(1,1) & 0.24973 $\pm$ 0.00017 &0.24976 \\
Beta(2,2) & 0.16661 $\pm$ 0.00012 & 0.16650 \\
Beta($\frac{1}{2}$,$\frac{1}{2}$) & 0.33310 $\pm$ 0.00021 & 0.33301 \\
Gamma(1,1) & 0.49936 $\pm$ 0.00031 & 0.49951 \\
Gamma(1,5) & 0.49945 $\pm$  0.00031 & 0.49951 \\
Gamma(5,1) & 0.16653 $\pm$  0.00013 & 0.16650  \\ \hline
		\end{tabular}
		\caption{Empirical and theoretical values of the expected fraction of energy loss, $\langle L \rangle$, for different initial velocity distributions, and considering a system of $N=10000$ initial particles.}
	\end{center}
\end{table}

\subsection{Explosion-like inital condition}
In this section,  the same system of particles will be studied, however, a different initial condition will be applied. This initial condition will consider particles placed at negative initial positions starting with negative random velocities, and particles positioned at positive initial positions starting with positive random velocities.  This initial condition mimics an explosion at the origin.

Since particles with positive initial positions will never interact with particles on the left, the result of this particular ``explosive'' initial condition is the superposition of the results of two independent systems: the right and the left one. Therefore, the expected number of final particles in this condition, $\langle \tilde{X}^{expl}_N \rangle$, is only $2\langle \tilde{X}_{N/2} \rangle$ when half of the particles are on each side. For a random distribution of $N$ particles on two sides,
 $$\langle \tilde{X}^{expl}_N \rangle = \overset{N}{\underset{k=0}{\sum}}\mathbb{P}(N_{\leftarrow}=k)(\langle \tilde{X}_{k} \rangle +\langle \tilde{X}_{N-k} \rangle),$$
 where $\mathbb{P}(N_{\leftarrow}=k)$ is the probability that $k$ of the $N$ initial particles start on the left, and  $\langle \tilde{X}_{1} \rangle=1$, and $\langle \tilde{X}_{0} \rangle=0$. Since both sides are independent, the variance is the sum of the variances on each side. Moreover, the distribution of $\tilde{X}^{expl}_N$ for a random sorting of $N$ particles on each side of the axis will verify,
 $$\mathbb{P}( \tilde{X}^{exp}_N=j)= \overset{N}{\underset{k=0}{\sum}}\overset{j}{\underset{l=0}{\sum}}
 \mathbb{P}(N_{\leftarrow}=k) \mathbb{P}(\tilde{X}_{k}=l)\mathbb{P}(\tilde{X}_{N-k}=j-l).$$

\section{Conclusions}
A 1D particle system of $N$ identical point particles undergoing perfectly plastic collisions has been studied. The general goal was to understand the final configuration of the merged particles for all values of $N$, not just for the large $N$ limit.

The strategy used for dealing with this complex model has been to introduce a simpler new model, the aforemention non-physical model, solve it, and apply what was learned from this simpler model to the physical model. For the non-physical model, the distribution of the number of final particles, as well as its mean and variance, were explicitly calculated.  Using the non-physical model as inspiration, a procedure for the physical model was subsequently developed (Theorem 2), in which the number of final particles and their masses could be determined without the need to evolve the system. The resulting procedure is significant, and holds particular value when it comes to studying very large systems.   Also, it has been proven that the initial positions distribution does not play any role in determining the final configuration (Theorem 1). Furthermore, numerical evidence was presented to show that the initial velocity distribution has no effect on the final configuration.
This suggests that the random variable number of final particles is universal or distribution-free  (Conjecture 1). This observed universal behavior, although not yet formally proven, is noteworthy because it indicates that the system is robust and consistent across different initial conditions.
 Moreover, numerical evidence has been presented which support that the distribution of the final number of particles and their masses is the same in both physical and non-physical models (Conjecture 2). This surprising result strongly justifies the prior decision to introduce the non-physical model as a tool to study the physical one.

Additionally, numerical evidence was provided to understand the distribution of masses.  Specifically, a randomly selected final particle appears to have a fraction of the total mass that follows a power-law distribution with a probability density proportional to $1/s$ for $0<s<1$ (eq.~\ref{powerlaw}). Also shown is that the energy loss, unlike the previous cases, is significantly influenced by the initial velocity distribution. In this case, a highly accurate theoretical approximation for this loss is presented. Moreover, for large $N$, this energy loss only depends on the first two moments of the velocity distribution (eq.\ref{lost}).  Finally,  the manner in which the number of final particles changes for an explosive-like initial condition was analyzed.

Both conjectures merit further investigation. While finding proof for Conjecture 2 would imply that Conjecture 1 is true, there is something to be gained by testing Conjecture 1 independently. Both conjectures look into very different aspects of the model. The first one states that the initial position and velocity distributions, which describe the peculiarities at the beginning of the process, appear to be irrelevant for the final configuration (Conjecture 1). This is a great advantage, because if the evolution of an N-body system depended on these distributions, it would be extremely difficult to draw conclusions from a single realization. To provide a rough analogy, if the initial condition distributions were relevant, it would be almost impossible to make any meaningful statements about the Big Bang solely by observing the present universe.

On the other hand, Conjecture 2 will deal with another type of universality, which is equally interesting and more useful for understanding the model. This is the universality is related to the results that models produce. Models are known to be primarily defined by the way in which their constituent elements interact. As of yet, there is no general procedure for understanding whether two models with different interactions will manifest the same observables.
In our case, while most realisations under the same initial conditions yield different results (see Fig. 1), surprisingly the results obtained for the non-physical systems seems to be \textit{statistically equivalent} to the physical ones (Conjecture 2). This is similar to the universality classes of systems in the critical regime of phase transitions, where different models behave in the same way~\cite{libro_criticalidad1,universality1}. That rises the question: are there ``universality'' classes for non-equilibrium systems like the one presented in this paper? Results show that both, the physical and the non-physical model, behave in the same way.  This is compatible with a vision that proposes that both models belong to the same model category. Admittedly, looking into the existence of classes of mathematical models for non-equilibrium systems lacking a phase transition is not only fascinating, but it could also aid in better understanding non-equilibrium systems.

Finally, it is likely there are some challenges in trying to extend the system presented here to larger dimensions. In this case, it is necessary to adapt the point particles to finite particles in order for collisions to happen, and of course, it is also necessary to modify the initial conditions. It is expected for the number of final particles ($\tilde{X}_N$) to increase along with the dimension of space, as well as for the distribution of the final particle masses to change; however, the decreasing monotonic behavior of the probability density is expected to stay the same.
 In order to properly calculate these statistical properties it is fundamental to consider a key variable: the percentage of the total particles that can be considered at the surface (or that belongs to the ``propagating wavefront''). Typically, superficial particles are likely to ``escape'', experiencing a small number of plastic collisions in the process. On the other hand, those that start closer to the center will suffer significantly more collisions, becoming considerably massive particles. Therefore, as a general rule, for non-explosive random initial conditions,  one would say that the final particles that are farther away from the starting point will most likely be lower-mass particles.
  Conversely, under explosive initial conditions, the lower mass particles are likely to be both those farther away and those closer to the starting point.

   The concepts and outcomes presented in this paper might prove valuable in understanding and addressing other non-equilibrium processes. To provide a rough analogy,  the prior discussion on non-explosive initial conditions shares some similarities with the Big Bang expansion. As first order, it behaves like an out-of-equilibrium gas of interacting elemental particles that finally produces clusters of particles (atoms/stars/galaxies). In contrast, the example involving explosive initial conditions can be related to a supernova explosion~\cite{supernova} and the subsequent formation of stars.  In this analogy, plastic collisions are to final particles what gravity is to galaxies: just like stars and interstellar matter are bound together by gravity, original particles are bound by plastic collisions.

   In this context, it's also worth mentioning that the measurements and theoretical calculations of the mass distribution of stars and galaxies remain an active area of research~\cite{galaxy1,galaxy11,galaxy2,galaxy3,galaxy4}. Some keywords include the Initial Mass Function~\cite{galaxy0} (IMF) and the Present-Day Mass Function~\cite{galaxy11} (PDMF). The IMF describes the distribution of stellar masses at birth, while the PDMF describes the current distribution of masses.  The difference between these two densities is that the stellar mass distribution changes over time due to various dynamical phenomena, such as the depletion of low-mass stars through evaporation. An interesting observation is the predominance of small stars over large ones in both distributions, since the distribution is well-described mathematically by a power-law, or at least a power-law tail provides a good fit. This distribution is consistent with the mass distribution identified in our simple model, although the power exponent is different (here equal  to 1, see eq. 11). However, it is important to emphasise that this analogy is only a rough approximation.  Various physical phenomena in both scenarios limit the applicability of this analogy~\footnote{For example, galaxies are dynamic entities that will eventually collapse under gravitational forces. Thus, the analogy may be most relevant during a quasi-stationary period when galaxies remain relatively unchanged before their eventual collapse.  In the context of star formation, the analogy also has limitations; the mass of stars cannot exceed a certain limit, or they will collapse into a black hole.}.

   Nevertheless,
   from a modeling perspective, the findings outlined in this study could potentially pave the way for the development of more intricate models that closely mirror real-world scenarios.

\section*{ACKNOWLEDGMENT}
I would like to thank J. Porto Alonzo for her invaluable assistance in improving the clarity and readability of the manuscript. 

\section*{Appendix A:  Proof of Theorem 1}\label{appA}
Suppose $N$ particles start at positions $Y_1<Y_2< \dots <Y_N$ and with velocities $V_1,V_2, \dots, V_N$ respectively.  If we evolve the system, we end up with $\tilde{X}_{N}$  particles with masses $\mathbb{M}=(m_1,m_2,\dots, m_{\tilde{X}_N})$ and final velocities ($v^f_1, v^f_2,\dots, v^f_{\tilde{X}_N}$) with $v^f_i=\overline{V}_{M_{i-1}+1,M_i}$ with $M_k=\overset{k}{\underset{i=1}{\sum}} m_i$ for $k \in \Theta_{\tilde{X}_N}$ and $M_0=0$.

Final particles verify: $$\overline{V}_{1,m_1}<  \overline{V}_{M_1+1,M_2}< \overline{V}_{M_2+1,M_3}<\dots <\overline{V}_{\tilde{X}_N-1,M_{\tilde{X}_N}}.$$
In addition, every fused final particle verifies:
\begin{equation}\label{posi}
 \overline{V}_{M_{k-1}+1,M_k-1}>  \overline{V}_{M_{k},M_k}.
 \end{equation}
Otherwise, the last particle composing the final particle would not merge. The same argument holds for the last $j$ particles of the final particle, i.e.
 \begin{equation}\label{posi2}
 \overline{V}_{M_{k-1}+1,M_k-j}>  \overline{V}_{M_{k}-j+1,M_k}.
 \end{equation}

 Now we study the system, but in this case, starting from different positions $Y^{new}_1<Y^{new}_2< \dots <Y^{new}_N$, while maintaining the previous velocities. By evolving the system, we end up with $\tilde{X}^{new}_{N}$ particles with masses $\mathbb{M}^{new}=(m^{new}_1,m^{new}_2,\dots, m^{new}_{\tilde{X}^{new}_{N}})$.
 The equation~\ref{posi2} is again fulfilled, but $M_k$ is replaced by $M^{new}_k:=\overset{k}{\underset{i=1}{\sum}} m^{new}_k$,
    \begin{equation}\label{posi3}
 \overline{V}_{M^{new}_{k-1}+1,M^{new}_k-j}>  \overline{V}_{M^{new}_{k}-j+1,M^{new}_k}.
 \end{equation}

 Suppose the final configurations are different, $\mathbb{M}\neq \mathbb{M}^{new}$. Then there is a first final particle in which they are certain to differ.
   Without loss of generality, let us assume that this difference occurs in the first final particle, i.e. $m_1\neq m_1^{new}$.

Suppose first that $m^{new}_1$ is less than $m_1$, specifically $m^{new}_1 = m_1 - j$ with $j \in \Theta_{m_1-1}$. First note, the first final particle will always have a lower velocity than the second final particle,
 \begin{equation}
\overline{V}_{1,m^{new}_1}<\overline{V}_{m^{new}_1+1,m^{new}_1+m^{new}_2}.
 \end{equation}
 According to eq~\ref{posi3}, we have that $\overline{V}_{m^{new}_1+1,m^{new}_1+m^{new}_2}< \overline{V}_{m^{new}_1+1,m^{new}_1+j}$.
 So by putting this information together, it is verified:
 \begin{equation}\label{contra1}
\overline{V}_{1,m^{new}_1}< \overline{V}_{m^{new}_1+1,m^{new}_1+j}.
 \end{equation}
 However, if we look at the original system with the initial positions $Y_1, Y_2, \dots, Y_N$, we see that due to the equation~\ref{posi2}, it is fulfilled:
  \begin{equation}\label{contra2}
\overline{V}_{1,m^{new}_1}=\overline{V}_{1,m_1-j}> \overline{V}_{m_1-j+1,m_1}=\overline{V}_{m^{new}_1+1,m^{new}_1+j}.
 \end{equation}
 This last equation contradicts equation~\ref{contra1}, so $m^{new}_1$ cannot be less than $m_1$.

 In the same way, we can prove that $m^{new}_1$ cannot be greater than $m_1$. So $\mathbb{M}=\mathbb{M}^{new}$.

\section*{Appendix B: Proof of Theorem 2}\label{appB}
The process begins with a configuration of velocities $v_1 := v_i = (V_1, V_2, \dots, V_N)$ and a set of positions $Y_1 < Y_2 < \dots < Y_N$. The specific positions determine the sequence of collisions, dictating which ones occur initially and which ones follow. Importantly, this sequence doesn't impact the total number of final particles or the mass of each individual particle. For a detailed proof, refer to the end of this appendix. Consequently, in the following, we will assume a particular order in which the collisions take place

For particles 1 and 2 to merge, one of two alternatives must occur: (A)  $V_1>  V_2$, or (B) $V_1>  \frac{1}{s-1} \overset{s}{\underset{i=2} {\sum}} V_i$ for some $s>2$ (i.e. particle 1 collide with a fused particle that contains particle 2).   Alternatively, we can express that particle 1 merges with other particles if $s_1(v_i)\neq \emptyset$, where
 $$s_1(v_1)=\min\{s \in \Theta_{2,N}:  V_1> \overline{V}_{2,s} \},$$
with $\overline{V}_{j,k}=\frac{1}{k-j}\overset{k}{\underset{i=j} {\sum}} V_i$. If $s_1(v_1)\neq \emptyset$, particle 1 will collide with a merged particle comprising particles $2, 3, \dots,  s_1(v_1)$. In essence, a merged particle will form, incorporating particles  $1, 2 , \dots, s_1(v_1)$, and it will have a velocity  $\overline{V}_{1,s_1(v_1)}$.

Now, with this new merged particle, we work as we did before, i.e. as if it were the original particle 1, which can be merged when $s_2(v_1)\neq \emptyset$, where
 $$s_2(v_1)=\min\{s \in \Theta_{s_1+1,N}:  \overline{V}_{1,s_1}> \overline{V}_{s_1+1,s} \}.$$
 If $s_2(v_1) \neq \emptyset$, then the merged particle containing particles $1, 2, \dots, s_1(v_1)$ will inevitably collide with another merged particle containing particles $s_1(v_1)+1, 3, \dots, s_2(v_1)$. Consequently, a new merged particle will form, incorporating particles $1, 2, \dots, s_2(v_1)$, and it will possess a velocity of $\overline{V}_{1,s_2(v_1)}$. This process repeats until, for the first time,
$$\tilde{k}(v_1)=\min\{k \in \Theta_N: s_k(v_1)=\emptyset \ or \ s_k(v_1)=N \}.$$
with $$s_k(v_1)=\min\{s \in \Theta_{s_{k-1}(v_1)+1,N}:  \overline{V}_{s_0(v_1),s_{k-1}(v_1)}> \overline{V}_{s_{k-1}(v_1)+1,s} \},$$
and $s_0(v_1)=1$.

Finally, the resulting final particle 1, the leftmost particle, will be a fusion of particles $1,2, \dots, s_{\tilde{k}(v_1)}(v_1)$. That is, the mass will be $s_{\tilde{k}(v_1)}(v_1)$, and if its mass is smaller than $N$, the following condition will be satisfied~\footnote{Eq. 18 is equivalent to:  $$\overline{V}_{1,s_{\tilde{k}}(v_1)}< \min\{\overline{V}_{s_{\tilde{k}(v_1)}(v_1)+1,s_{\tilde{k}(v_1)}(v_1)+1},\dots, \overline{V}_{s_{\tilde{k}(v_1)}(v_1)+1,N}\}$$. }:
\begin{equation}\label{proof1}
\overline{V}_{1,s_{\tilde{k}(v_1)}(v_1)}< \overline{V}_{s_{\tilde{k}(v_1)}(v_1)+1,s} \quad \forall s \in \Theta_{s_{\tilde{k}(v_1)}+1,N}.
\end{equation}
 The general case of $s_{\tilde{k}(v_1)}(v_1)$,  which includes the value $N$, can be written as:
\begin{equation}\label{sinevol}
\begin{split}
s_{\tilde{k}(v_1)}&(v_1)=\tilde{m}(v_1)\\
&=\min\{j \in \Theta_{1,N}: \overline{V}_{1,j}<  \overline{V}_{j+1,i} \quad \forall i \in \Theta_{j+1,N+1}\}.
\end{split}
\end{equation}
 Where, to improve the notation, an additional ``phantom'' particle is added; particle $N+1$, positioned to the right of particle $N$, with a velocity equal to $2V_{\text{max}} := 2\max\{V_1, V_2, \dots, V_N\}$. The upper row in Fig. S1 presents an illustration of a system involving $N=9$ particles, solved through the temporal evolution of the process. In contrast, the analogous scenario without temporal evolution, as outlined by equation~\ref{sinevol}, is portrayed in the upper row of Fig. S2.
\begin{figure}
\raggedright
 \includegraphics[width=8cm]{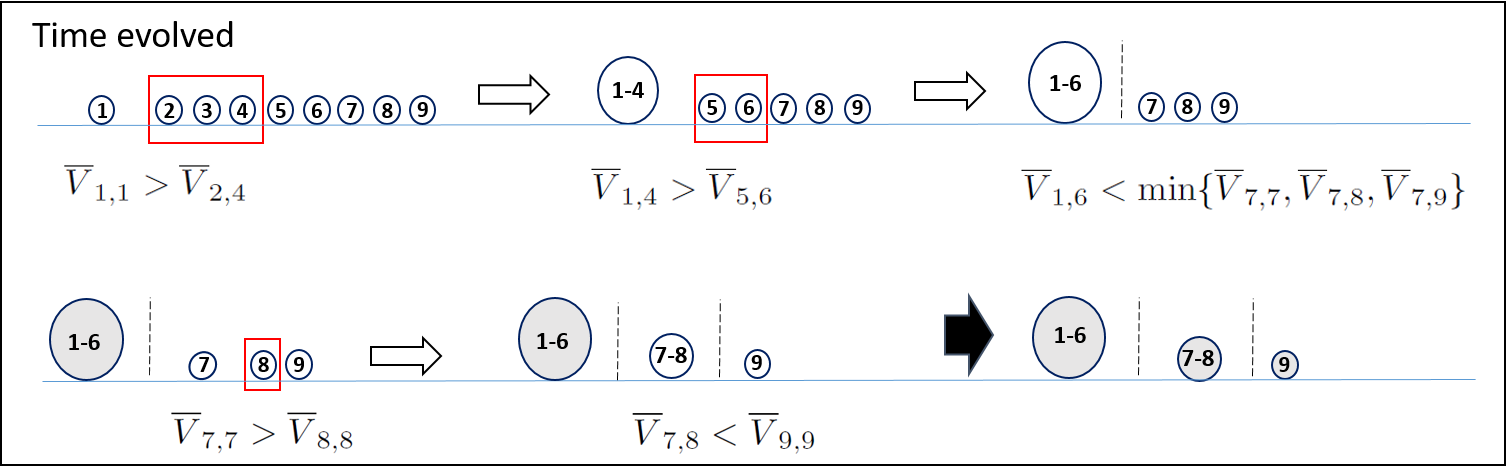}
\textbf{Fig. S1.} Solution based on system evolution.\label{fig4}
\end{figure}

Now that we understand the composition of the final particle 1, comprising particles $1, 2, \dots, \tilde{m}(v_1)$, we can apply a similar analysis to the remaining particles on the right-hand side. Assuming $\tilde{m}(v_1) < N$, we initiate the analysis starting with particle $\tilde{m}(v_1)+1$ to investigate potential mergers with the remaining particles.
Let $v_2=v_1[\tilde{m}(v_1)+1:N]$=($V_{\tilde{m}(v_1)+1},V_{\tilde{m}(v_1)+2},\dots, V_N$) and define
 $$s_1(v_2)=\min\{s \in\Theta_{\tilde{m}(v_1)+2,N}:  V_{\tilde{m}(v_1)+1}> \overline{V}_{\tilde{m}(v_1)+2,s} \}.$$
If $s_1(v_2) \neq \emptyset$, particle $\tilde{m}(v_1)+1$ will collide with a merged particle comprising particles $\tilde{m}(v_1)+2, \tilde{m}(v_1)+3, \dots, s_1(v_2)$. A merged particle will form, incorporating particles $\tilde{m}(v_1)+1, \tilde{m}(v_1)+2, \dots, s_1(v_2)$, and it will have a velocity $\overline{V}_{\tilde{m}(v_1)+1, s_1(v_2)}$. This fused particle can fuse with other particles if $s_2(v_2) \neq \emptyset$, with
 $$s_2(v_2)=\min\{s \in \Theta_{s_1(v_2)+2,N}:  \overline{V}_{s_1(v_2)+1,s}> \overline{V}_{s_1(v_2)+2,s} \}.$$
 If $s_2(v_2) \neq \emptyset$, the fused particle will collide. The new fused particle will be formed by particles $s_1(v_1)+1, s_1(v_1)+2, \dots, s_2(v_2)$, and this process continues until the first time that
 $$\tilde{k}(v_2)=\min\{k \in \Theta_N: s_k(v_2)=\emptyset \ or \ s_k(v_2)=N \}.$$
with $$s_k(v_2)=\min\{s \in \Theta_{s_{k-1}(v_2)+1,N}:  \overline{V}_{s_0(v_2),s_{k-1}(v_2)}> \overline{V}_{s_{k-1}(v_2)+1,s} \},$$
and $s_0(v_2)=s_{\tilde{k}(v_1)}(v_1)+1$. It is crucial to reemphasize that, within the context of this proof,
$$s_{\tilde{k}(v_2)}(v_2)=\tilde{m}(v_2).$$
This continues until we have $\tilde{j}$ final particles, with
 $$\tilde{j}= \min\{j \in \Theta_N: \overset{j}{\underset{h=1}{\sum}} \tilde{k}(v_h)=N\}.$$
\begin{figure}
\raggedright
\includegraphics[width=8cm]{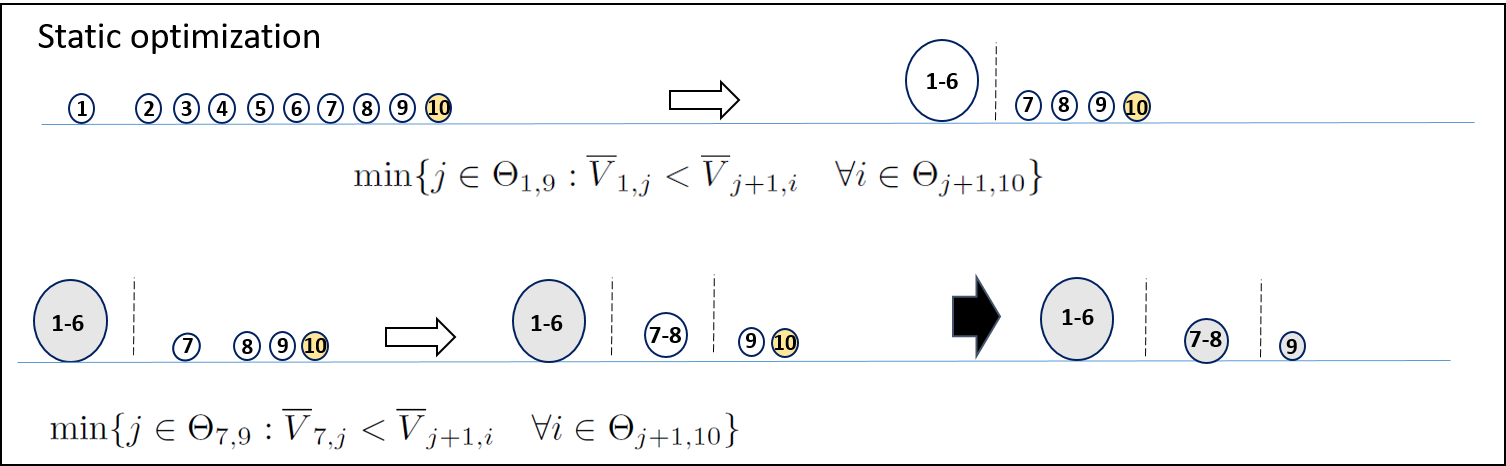}
\textbf{Fig. S2.} Solution without relying on system evolution.\label{fig5}
\end{figure}
Finally, evolving the system, we obtain the following final masses:
\begin{equation*}
\begin{split}
\mathbb{M}&=(s_{\tilde{k}(v_1)}(v_1),s_{\tilde{k}(v_2)}(v_2),\dots,s_{\tilde{k}(v_{\tilde{j}})}(v_{\tilde{j}})),\\
&=(\tilde{m}(v_1),\tilde{m}(v_2),\dots,\tilde{m}(v_{\tilde{j}})), \\
&=(\tilde{m}(G^0(v_i)),\tilde{m}(G^1(v_i)),\dots, \tilde{m}(G^{\tilde{X}_N(v_i)-1}(v_i))).
\end{split}
\end{equation*}
 To conclude the proof, note that the last equality is valid, since $v_j=G^{j-1}(v_i)$ and
 \begin{equation*}
\begin{split}
 \tilde{X}_N(v_i)&=min\{k\in \Theta_{N}: G^{k}(v_i)=\emptyset\},\\
 &=min\{k\in \Theta_{N}: \overset{k-1}{\underset{j=0}{\sum}}\tilde{m}(G^j(v_i))=N\}.
 \end{split}
\end{equation*}

\section*{Appendix C: Algorithm for computing $\tilde{X}_N$ and $\mathbb{M}$.}\label{appC}
Based on Theorem 1, a simple algorithm for computing $\tilde{X}_N$ and $\mathbb{M}$ is presented here. For clarity, the main algorithm is first introduced, followed by an auxiliary function, $\tilde{m}$, which is then explicitly detailed as part of the main algorithm.

\begin{algorithm}
\SetAlgoLined
v[1:$N$] $\leftarrow$ randomF($N$);  \# N velocities

v[$N$+1] $\leftarrow 2$ $\cdot$ max(v); \# auxilar velocity

vi $\leftarrow$ v ;

w $\leftarrow$ 0 ; s $\leftarrow $1

$\mathbb{M}$ $\leftarrow$ vector();

\While{w $< N$} {

$\mathbb{M}$[s] $\leftarrow$ $\tilde{m}$(vi) ;

w $\leftarrow$ masses[s]+w ;

vi $\leftarrow$ v[(w+1):length(v)] ;

s $\leftarrow$ s+1 ;

}

$\tilde{X}_N$ $\leftarrow$ length($\mathbb{M}$) ; \# number of final particles

\caption{Main algorithm for $\tilde{X}_N$  and $\mathbb{M}$.}
\end{algorithm}

\begin{algorithm}
\SetAlgoLined
$\tilde{m}$=function(v)\{

$N \leftarrow$ length(v)-1

$\Theta$ $\leftarrow 1:N$; $\Lambda$ $\leftarrow 2:(N+1)$;

k  $\leftarrow$1; i $\leftarrow$ 1;

\While{$\Lambda[i]$ $<$ N+1}{

i $\leftarrow$ 1;

$\Lambda$ $\leftarrow$ ($\Theta$[k]+1):$(N+1)$ ;

vl $\leftarrow$ mean(v[1:$\Theta$[k]]);

vr  $\leftarrow$ mean(v[$\Lambda$[1]:$\Lambda$[i]]);

k $\leftarrow$ k+1;

\While{vl$<$vr \& $\Lambda$[i] $<$N+1}{

i $\leftarrow$ i+1 ;

vr $\leftarrow$ mean(v[$\Lambda$[1]:$\Lambda$[i]]);

}

}
mass $\leftarrow$ k-1;

return(mass)
\}

\caption{$\tilde{m}$ function.}
\end{algorithm}

\section*{Appendix D: Evidence for Conjecture 1}\label{appD}
It has been proved that $\tilde{X}_2$ snf $\tilde{X}_3$ do not depend on the initial velocity distribution (they are universal).  The number of final particles,  $\tilde{X}_N$, was studied numerically for different initial velocity distributions
 $F_1=$Uniform(-1,1),  $F_2=$Normal(0,1),  $F_3=$Normal(10,1),  $F_4=$Exponential(1),  $F_5=$Exponential(10),  $F_6=$Gamma(1,10),  $F_7=$Gamma(1,0.1),  $F_8=$Beta(1,1),  $F_9=$Beta(2,2), $F_{10}=$Beta($\frac{1}{2}$,$\frac{1}{2}$).

  Let $F_{n,k}$ be the empirical distribution of $\tilde{X}_N$ when the initial velocity distribution is $F_k$. Ir order to test if $\tilde{X}_N$ has the same distribution across the ten studied distributions, the KS statistic is computed between each pair for comparison,
  $$D_{j,k}=\underset{x}{sup}|F_{n,j}(x)-F_{n,k}(x)|,$$
   and then the maximum value is determined,
$$D_{max}=max\{D_{1,2}, D_{1,3},\dots,D_{1,10},D_{2,3},D_{2,4}\dots,D_{2,10},\dots, D_{9,10}\}.$$
\begin{figure}
\raggedright
 \includegraphics[width=8cm]{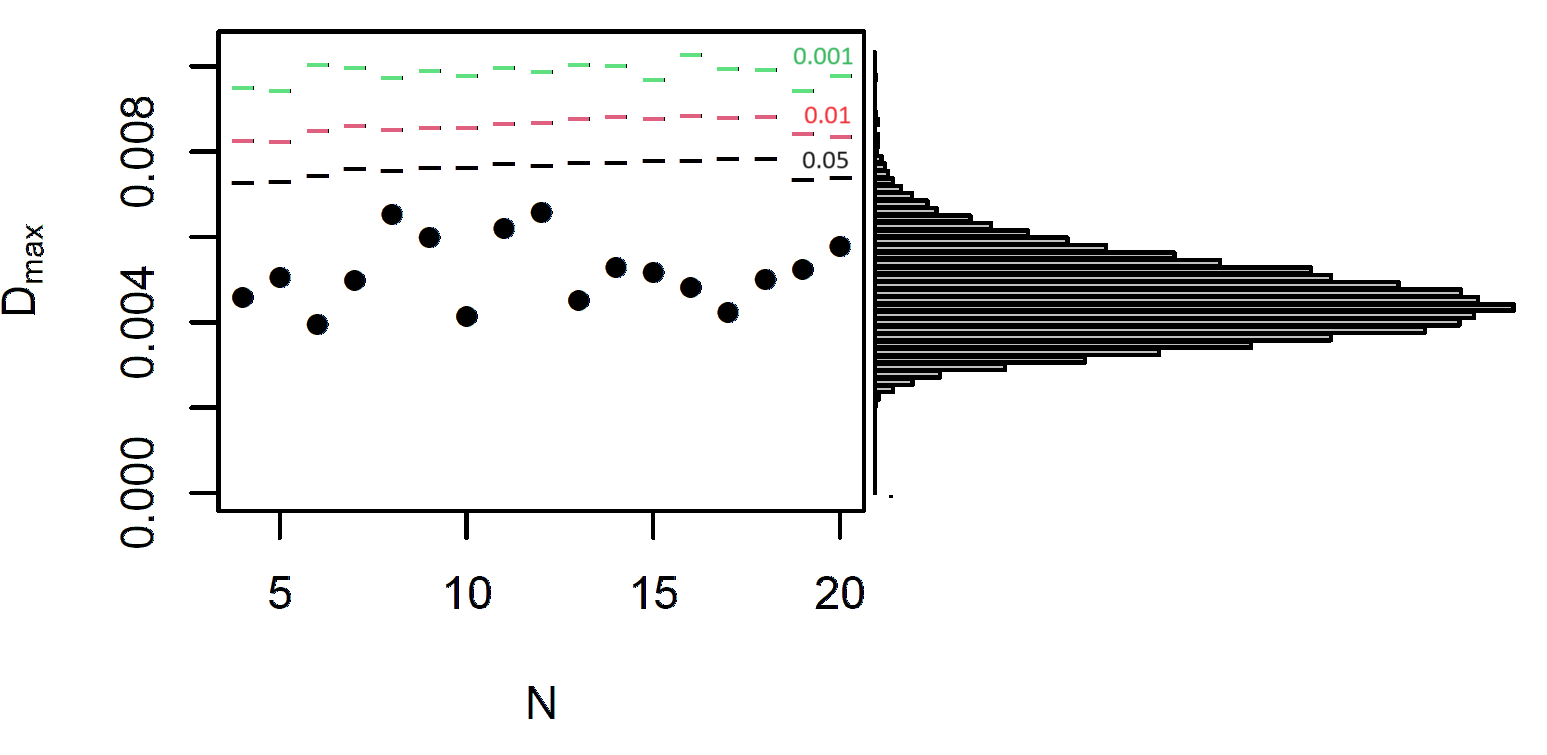}

\textbf{Fig. S3.} Black points corresponds to $D_{max}$ as a function of $N$. Segments represent the empirical quantiles 0.95 (black), 0.99 (red) and 0.999 (green) of the statistic $D^{null}_{max}$ under the hypothesis of equal distribution. On the left the empirical distribution of $D^{null}_{max}$ for $N=20$. Similar histograms are obtained for the different values $N$.
 \end{figure}
 Fig. S3 shows the statistics $D_{max}$ for $N=4,5,6,\dots, 20$ (black points). In addition to these data, empirical threshold values for significance levels of 0.05, 0.01, and 0.001 are also shown (color segments). To determine these threshold values, simulations were conducted under the null hypothesis of equal distribution for each value of $N$.  The distribution of $D_{max}$ under the null hypothesis is depicted on the right for the case $N=20$. Similar histograms are obtained for different values of $N$.

  To determine $D_{max}$ under the null hypothesis, denoted from now on as $D^{null}_{max}$, first
  $$D^{null}_{j,k}=\underset{x}{sup}|F^{null}_{n,j}(x)-F^{null}_{n,k}(x)|,$$
  is computed. Where $F^{null}_{n,j}(x)$ is the $j$ estimation (or equivalent a random realization $j$) of the empirical distribution of $\tilde{X}_N$ when considering the initial velocity distribution to follow the law $F^{null}(x)=\frac{1}{10}\overset{10}{\underset{i=1}{\sum}} F_i(x)$. Finally,
  $$D^{null}_{max}=max\{D^{null}_{1,2}, D^{null}_{1,3},\dots,D^{null}_{1,10},D^{null}_{2,3},\dots, D^{null}_{9,10}\},$$
is computed. For a given $N$, this procedure is repeated 100000 to obtain a good description of the  $D^{null}_{max}$ distribution. Finally the quantiles 0.05, 0.01, and 0.001 are estimated and presented in the figure.  As observed in Fig. S3, none of the observed statistics $D_{max}$ exceed the 0.05 threshold, i.e. the $D_{max}$ values observed are compatible with those generated by $D^{null}_{max}$.  In summary, the results presented here, along with those shown in the main text, strongly support Conjecture 1.

\section*{Appendix E: Evidence for Conjecture 2.}\label{appE}
In Table 1, evidence was presented supporting the claim that the mean of $\tilde{X}N$ is equal to that of $\tilde{Z}N$. Here, we analyze the variance, providing evidence that both random variables also share the same variance. The variance of $\tilde{Z}N$ is given by eq.~\ref{varf}, while for the physical model, the variance is estimated through simulations. In this case, a 95\% confidence interval is presented, given by [$(n-1)s^2/\chi^2{0.025,n-1}, (n-1)s^2/\chi^2{0.975,n-1}$], where $\chi^2{\alpha,n-1}$ corresponds to the $\alpha$ quantile of a Chi-squared distribution with $n-1$ degrees of freedom. Here, $n=30000$ was defined as the number of simulations. The results are shown in the following table. As can be observed, the variance of $\tilde{Z}_N$ falls within the confidence interval for the variance of $\tilde{X}_N$. This provides strong evidence in favor of both random variables sharing the same variance.

\begin{table}[h!]
	\begin{center}
		\begin{tabular}{|c|r|l|} \hline

   N & \quad  \quad  Var($\tilde{Z}_N$) \quad  \quad \quad  \quad &\quad  \quad  Var($\tilde{X}_N$)  \\ \hline
    2  &     $ \frac{1}{4}$\quad \quad \quad \quad   & \quad
    \quad \quad $\frac{1}{4}$  \\
    3  &      $ \frac{17}{36}$\quad \quad \quad \quad  & \quad
    \quad \quad $\frac{17}{36}$  \\
    4  &   $ \frac{95}{144}\approx  0.6597 $ &     [0.6532, 0.6745] \\  
    5  & $   \frac{274}{120}\approx  0.8197 $&  [0.8044, 0.8306] \\ 
    6  & $ \frac{3451}{3600}\approx  0.9586$ &    [0.9437, 0.9744] \\
    7  & $ \frac{190699} {176400}\approx  1.0811$  &    [1.0555, 1.0898]\\ 
    8  &$ \frac{839971}{705600}\approx  1.1904 $ &  [1.1580, 1.1956] \\
    9  &$ \frac{8186939}{6350400} \approx  1.2892$  &  [1.2807, 1.3223]  \\
   10 &$ \frac{350339}{254016} \approx  1.3792 $ &  [1.3589, 1.4031] \\ \hline 
		\end{tabular}
\end{center}
\raggedright
\textbf{Table S1.} Exact Var($\tilde{Z}_N$) and approximate Var($\tilde{X}_N$) for different values of $N$. The approximate data correspond to the 95\% confidence interval obtained from n=30000 simulations. For $N=2$ and 3 only exact results are presented.
\end{table}

Now, evidence is presented to show that $\tilde{X}_N$ and $\tilde{Z}_N$ share the same probability law, i.e., $\mathbb{P}(\tilde{X}_N=k)=\mathbb{P}(\tilde{Z}_N=k)$ for all $k\in \Theta_N$. For $N=2$ and $N=3$, this has already been shown; for $N$ greater than 3, results from simulations will be presented.
To quantify the similarity between both distributions, the empirical distribution of $\tilde{X}_N$, denoted as $F_n(x)$, will be compared to the theoretical distribution of $\tilde{Z}_N$, denoted as $F_{theor}(x)$, using the Kolmogorov-Smirnov statistic. The theoretical distribution, defined as $F_{theor}(x)=\mathbb{P}(\tilde{Z}_N\leq x)$, can be computed exactly from (Thm. 1) as
 $$F_{theor}(x)=\overset{x}{\underset{k=1}{\sum}}\frac{|c(N,k)|}{N!},$$
where $c(n,k)$ is the Stirling number of the first kind. To obtain $F_n(x)$, simulations for the physical model were performed, considering a random Uniform(-1,1) initial velocity distribution (which is not relevant according to Conjecture 1), and applying the algorithm from Appendix C. Then finally, the KS statistic is computed,
$$D=\underset{x}{sup}|F_n(x)-F_{theor}(x)|.$$
\begin{figure}
\raggedright
 \includegraphics[width=8cm]{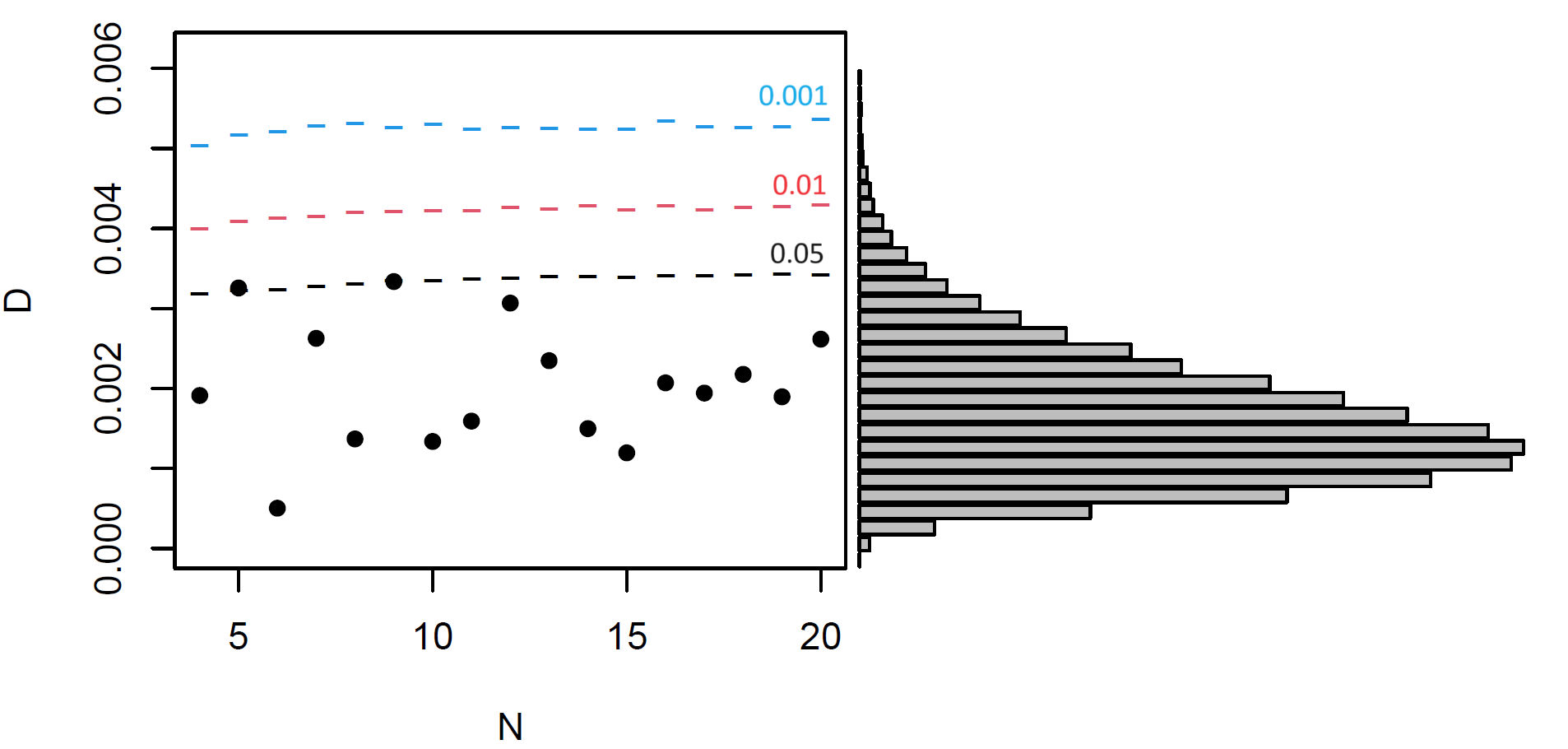}

\textbf{Fig. S4.} Black points corresponds to $D$ as a function of $N$. Segments represent the empirical quantiles 0.95 (black), 0.99 (red) and 0.999 (green) of the statistic $D$ under the hypothesis of equal distribution. On the left the empirical distribution of $D_{null}$ for $N=20$. Similar histograms are obtained for the different values $N$.
 \end{figure}
The statistic $D$ is calculated for $N=4,5,\dots, 20$. This data is represented as black points in Fig. S4. Alongside these values, threshold values for significance levels of 0.0 5, 0.01, and 0.001 are also shown. To obtain these threshold values, new simulations under the null hypothesis of equal distribution were conducted for each value of $N$.  This is necessary because the Kolmogorov-Smirnov statistic ($D$) is not distribution-free for discrete variables in our case.   The significance levels, represented as color segments, are shown for each value of $N$, considering 100,000 replicates. The distribution of $D$ under the null hypothesis is depicted on the right for the case $N=20$. Similar histograms are obtained for different values of $N$. Note  that almost all statistics have values smaller than the 0.05 threshold; only two have values close to this threshold. Under the null hypothesis, this is what is expected since 16 tests are performed. If corrections for multiple comparisons are applied, none of the statistics exceed the corrected threshold.


\section*{Appendix F: Conditional Mass distribution.}\label{appF}
In the main text the distribution of the size of random
Fig. S3  shows $\mathbb{P}(\tilde{S}_{N,n}>s)$ as a function of $s$ for $N=10000$ and three values of $n$.
\begin{figure}
  \raggedright
    \includegraphics[width=8cm]{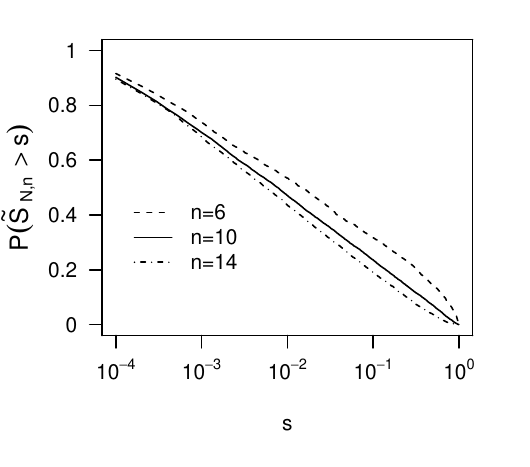}
\textbf{Fig. S4.} $\mathbb{P}(\tilde{S}_{N,n}>s)$ as a function of $s$ for $N=10000$ and $n=\{6,10,14\}$.
\end{figure}

\vspace{20cm}


\bibliographystyle{spbasic} 

\end{document}